\documentclass[aps, prx, reprint, superscriptaddress]{revtex4-2}
\usepackage{graphicx}  
\usepackage{dcolumn}   
\usepackage{bm}        
\usepackage{amssymb}   
\usepackage{color}
\usepackage{amsmath}
\usepackage{amsfonts, amsmath, amssymb}
\usepackage{appendix}
\usepackage{xcolor}
\begin{document}

\title{Analyzing Post-transcriptional Regulation in Stochastic Gene Expression \\ Models Using Partitioned Poisson Arrivals}

\author{Kenny Wong} 
\email{\texttt{wong.1216@osu.edu}}
\affiliation{Department of Physics, The Ohio State University, Columbus, Ohio 43210}
\affiliation{Department of Physics, University of Massachusetts Boston, Boston, MA 02125}

\author{Argenis Arriojas}
\affiliation{Center for Personalized Cancer Therapy, University of Massachusetts Boston, Boston, MA 02125}

\author{Sho Inaba}
\affiliation{Department of Physics, University of Massachusetts Boston, Boston, MA 02125}

\author{Hodjat Pendar}
\affiliation{Department of Mechanical Engineering, Virginia Tech, Blacksburg, VA 24061}

\author{Abhyudai Singh}
\affiliation{Department of Electrical \& Computer Engineering, University of Delaware, Newark, DE 19716}

\author{Rahul Kulkarni}
\email{\texttt{rahul.kulkarni@umb.edu}}
\affiliation{Department of Physics, University of Massachusetts Boston, Boston, MA 02125}

\date{\today}

\begin{abstract}
    Gene expression is a stochastic process that allows for fluctuations in protein levels {that can give rise to} phenotypic heterogeneity within a population of genetically identical cells. Thus, there is great interest in quantifying how natural variation (noise) in gene expression is impacted by cellular control mechanisms, such as the various mechanisms pertaining to post-transcriptional regulation. Although previous research has developed a general analytical framework to compute the exact moments of mRNA distributions for any promoter-based regulatory motif, and the exact mRNA distribution {itself} in some cases, a similar framework for protein fluctuations is currently lacking. Here, we invoke the partitioning property of Poisson arrivals to map a general class of stochastic models of post-transcriptional regulation onto models that resemble promoter-based regulation. This approach leads to {exact analytical results for the moments of protein distributions, and in certain cases the full distribution {itself}, using known exact results for mRNA distributions undergoing arbitrary promoter-based regulation.} We {further} extend the framework to incorporate transcriptional bursting, leading to a versatile, unifying analytical framework for analyzing post-transcriptional regulation in stochastic gene expression. 
\end{abstract}

\maketitle

\section{Introduction}
Even in isogenic cell populations, the intrinsically stochastic nature of gene expression can give rise to large fluctuations in mRNA and protein levels {that} create phenotypic heterogeneity \cite{Raj-Cell-2008, Horowitz-PhysBiol-2017}. In some cases, noise needs to be limited \cite{Raj-Cell-2008} because it can disrupt cellular function. On the other hand, noise can {also} be beneficial in {high-stress} environments and play a role in bet-hedging phenomena, driving probabilistic cell-fate outcomes such as drug resistance in melanoma \cite{Schuh-CellSys-2020, Shaffer-Nature-2017}, bacterial persistence \cite{Mirouze-plos-2011}, and {latent-to-active} switching in HIV-1 infections \cite{Weinberger-Cell-2005, Kumar-PRL-2014}. Thus, stochastic gene expression is a {recurring} theme in modern biological research{,} and developing analytical techniques to characterize noise in gene expression is crucial for a rigorous understanding of the underlying mechanisms that induce cell-to-cell variability in clonal cell populations. 

To this end, prior work has developed an analytical framework for computing the exact moments of mRNA distributions under arbitrary promoter regulation by representing mRNA production as a Markovian arrival process (MAP) and then analyzing the corresponding system of master equations \cite{Peccoud-TPB-1995,Sanchez-PNAS-2008,Herbach-SIAM-2019,Lippitt-SIAM-2022}. However, a comparable analytical framework for characterizing protein fluctuations in general models of \textit{post-transcriptional} regulation remains lacking. Many post-transcriptional processes can be described by finite networks in which individual transcripts transition among regulatory states with state-dependent translation and degradation rates \cite{Grandi-CellSys-2024}. At the protein level, the contributions from each transcript (that randomly arrives and dies) must then be integrated over their stochastic state trajectories. This makes the development of the corresponding analytical framework for the post-transcriptional level much more challenging than that for promoter-based regulation of mRNA production. The fact that post-transcriptional regulatory mechanisms are extremely diverse \cite{Furlan-BinB-2021} (e.g.{,} small RNA (sRNA), multi-step mRNA decay, nuclear export, alternative splicing) motivates the development of a general analytical framework that exploits the common mathematical structure of post-transcriptional networks, independent of any specific mechanism. 

Using the partitioning of Poisson arrivals (PPA) approach for analyzing stochastic models of gene expression developed in Ref. \cite{Pendar-PRE-2013}, we map {general} models of post-transcriptional regulation {with linear reaction propensities} onto reduced models {corresponding to} promoter-based regulation, allowing us to leverage results from Refs. \cite{Sanchez-PNAS-2008, Nossan-BJ-2024} for the post-transcriptional setting. {The key point is that protein production from an mRNA transcript transitioning through multiple states becomes, for a single partition, mathematically equivalent to mRNA production from a promoter switching among different regulatory states. We use this equivalence to obtain exact protein moments and also exact protein distributions in selected cases.} The combination of these approaches constitutes an analytical framework for analyzing both mRNA and protein distributions that arise from general models of post-transcriptional regulation. Additionally, we further extend the framework to analytically incorporate transcriptional bursting and demonstrate its versatility on various complex, biologically motivated post-transcriptional model schemes.

\section{Model \& PPA Mapping}
Consider the stochastic model of post-transcriptional regulation depicted in Figure 1A. A single DNA molecule produces mRNA strands at a rate of $k_m$, which are initially in state $M_1$. The mRNA can freely visit the $n$ available mRNA states{; in} a given state $i$, the mRNA strand produces proteins at a rate of $\kappa_i$ with a corresponding degradation rate of $\gamma_i$. mRNA transitions {from} arbitrary post-transcriptional regulatory {state} $j$ {to state} $i$ occur at rate $\alpha_{ij}$. These states change the ability for mRNA to produce proteins and/or the rate of degradation. Proteins degrade at a rate of $\mu_p$. All production, degradation, and state-to-state transition events are {assumed to be independent Markov counting processes. Throughout this paper, ``arbitrary post-transcriptional regulation'' refers to an arbitrary finite network of {first-order} transitions and degradation reactions.}  

\begin{figure*}
    \centering
    \includegraphics[width=\linewidth]{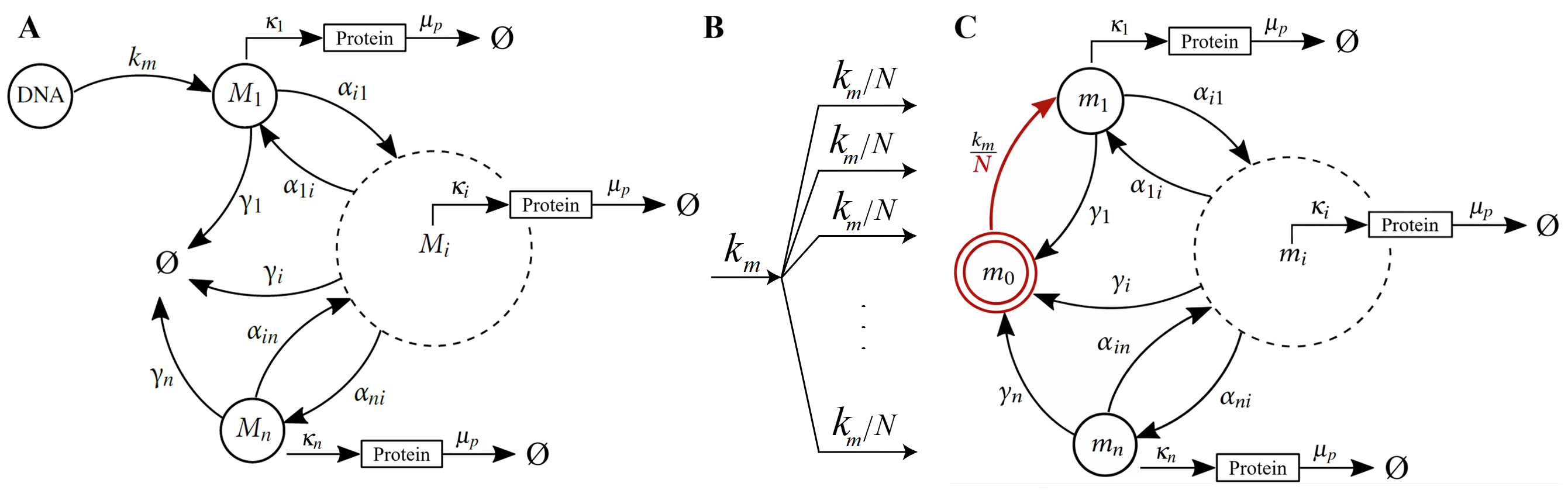}
    \caption{A) Representation of a general model of post-transcriptional regulation. mRNA molecules may be in several mRNA-states $i \in \{1, 2, ...,n\}$, with each state having distinct production and/or degradation rates{. In} state $i$, mRNA produce proteins at rate $\kappa_i$ and degrade at rate $\gamma_i$. We write $\alpha_{ij}$ as the transition rate between states $j$ and $i$. Proteins degrade at rate $\mu_p${.} B) mRNA Poisson arrivals partitioned into $N$ identical Poisson processes with a rescaled arrival rate of $k_m/N$; taken from {Ref.} \cite{Pendar-PRE-2013}. C) Reduced model of (A) constrained to either 0 or 1 mRNA when $N \rightarrow \infty$, resembling a promoter-based model. mRNA degradation is modeled as a transition back into the $m_0$ state.}
    \label{figenter-label}
\end{figure*}

{To simplify the analysis of the general model using the PPA mapping, let us begin by partitioning the mRNA arrivals into $N$ types (Figure 1B) with each unique type corresponding to an individual partition. Each mRNA arrival is randomly assigned to an arbitrary type $i$ drawn from $\{ 1, 2, \dots, N \}$ with uniform probability $1/N$. From the partitioning theorem of Poisson processes, the arrival processes of mRNAs for each type are each independent, identical Poisson process with a rescaled arrival rate $k_m/N$. Correspondingly, the number of arrivals up to time $t$ in each partition are independent, identically distributed random variables drawn from Poisson$(k_mt/N)$. Each partition independently contributes to the total number of mRNA arrivals 
\begin{equation}
    Y = \sum^N_{i=1} y_i  
\end{equation} where $y_i$ corresponds to number of mRNA arrivals of the $i$-th type. This property is central to the PPA mapping and also applies to the protein count. Thus, at any time $t$, we have a mapping between the dynamics of the original model and the dynamics of $N$ identical subsystems.}

{Now, in the limit $N \to \infty$, it was previously shown that the probability of each partition having more than one mRNA arrival can be neglected  \cite{Pendar-PRE-2013} because it is infinitely more likely that the additional arrivals get assigned to any other type with probability $\mathcal{O}(N^{-1})$ whereas where multiple arrivals have probability $\mathcal{O}(N^{-2})$. This constrains the mRNA count in each partition to be either 0 or 1. From here, the absence of the mRNA in each partition can be represented by the additional $m_0$ state (Figure 1C). Figure 1C is referred to as the \textit{reduced model} which arises only after taking $N\to\infty$ in the partitioning. As a result, the original model now maps onto a model that is mathematically equivalent to a model of promoter-based regulation where the mRNA states now resemble promoter states where $m_0$ is taken to be an additional inactive promoter state ({Figure} 1C).}


The reduced models are much easier to analyze, and since each partition independently contributes to the total protein and mRNA count, their probability distributions at an arbitrary time $t$, $\phi_q(x_q, t)$ with $q \in \{ 1, 2, \dots, N\}$, can be used to construct {the probability distribution} of the original model $\Phi(X,t)$ by
\begin{equation}
    \Phi(X,t) = \lim_{N \rightarrow \infty} \sum_{x_1 + x_2 \ldots + x_N = X} \prod_{q=1}^N \big[ \phi_q(x_q,t) \big].
\end{equation}
Recalling that the dynamics of each partition are all identical, the probability generating function for the reduced model $g(z,t) := \sum_{x=0}^\infty z^x \phi(x,t)$ (dropping the subscript) can be similarly used to construct {the generating function} of the original model $G(z,t) := \sum_{X=0}^\infty z^X \Phi(X,t) $ via 
\begin{equation}
    G(z,t) = \lim_{N \rightarrow \infty} \big[g(z,t)\big]^N.
\end{equation}
Working in the limit $N \to \infty$, {expressions to leading order in} $1/N$ are already exact. Consequently, we have previously shown \cite{Pendar-PRE-2013} that $g(z,t)$ can {alternatively be} related to $G(z,t)$ by 
\begin{equation}
    G(z,t) = \lim_{N \rightarrow \infty} \exp \{ N(g(z,t) -1) \}.
\end{equation}
In cases where the reduced model is still too complex for $g(z,t)$ to be analytically obtained, {Eqs.} (3) and (4) still relate the moments of the reduced model to the original model. Differentiating either equation with respect to $z$ {and evaluating at} $z=1$, it is straightforward to show that the mean and variance of the original model are connected to the reduced model by
\begin{align}
    \langle X \rangle  & = \lim_{N \rightarrow \infty} N \langle x \rangle, \\
    \langle X^2 \rangle - \langle X \rangle^2 &= \lim_{N \rightarrow \infty} N \big[\langle x^2 \rangle - \langle x \rangle^2 \big] 
\end{align}
with $\langle \cdot \rangle$ denoting the expected value operator. {Higher-order} moments can be found by successively differentiating the generating function further. The corresponding Fano Factor ($F$), typically used to quantify noise strength, is given by the variance upon the mean 
\begin{align}
    F := \frac{\langle X^2 \rangle - \langle X \rangle^2}{\langle X \rangle}
\end{align}
where it is easy to see that the reduced and original models have the same Fano factor. For Poisson arrivals, the Fano factor is exactly $1$; deviations from unity indicate deviations from the Poisson process \cite{Munsky-Science-2012}. {Noting that $\langle x \rangle$ scales as $\sim1/N$, the $N \langle x \rangle^2$ term vanishes in Eq. (6) in the limit $N\rightarrow\infty$.} Thus, the steady-state Fano factor (suppressing the time index) for the reduced model {can} be more conveniently expressed as
\begin{equation}
    F = \frac{\partial^2_z g(z)}{\partial_z g(z)}\bigg\vert_{z=1} + 1.
\end{equation}
In the following section, we flesh out the mathematical arguments for the PPA mapping for the general model depicted in Figure 1A. Note that the PPA mapping can be applied at both the mRNA and protein {levels}, as we will show.

\section{Leveraging Exact Results from Reduced Models}
\subsection{Exact mRNA Distributions}
We start {by} illustrating the PPA mapping by obtaining an exact steady-state mRNA distribution in the general model of post-transcriptional regulation. {The essential simplification due to the PPA mapping is that each reduced partition is mathematically equivalent to a promoter-state model}. Let us define the time-dependent distribution $P_t(\mathbf{\mathcal{M}})${,} where the vector $\mathbf{\mathcal{M}} = (m_1, m_2, ... , m_n)$ with $m_i$ denoting the amount of mRNA in state $i$. For convenience, we define the vectors corresponding to particular realizations of $\mathcal{M}$ as $\overrightarrow{m}_1=( 1, 0, 0, ...)$, $\overrightarrow{m}_2=( 0, 1, 0, ...)$, etc.{,} {such that $\overrightarrow{m}_i$ denotes a single mRNA molecule occupying state $i$. The set $\{ \overrightarrow{m}_i\}^n_{i=1}$ forms the standard basis of $\mathbb{R}^n$, and we additionally write $\overrightarrow{m}_0 = (0,0,\dots,0)$ for the empty (no-mRNA) configuration of the reduced model introduced below}. $P_t(\mathbf{\mathcal{M}})$ evolves according to the master equation
\begin{equation}
    \partial_t P_t(\mathbf{\mathcal{M}}) = \mathcal{D}_1[P_t(\mathbf{\mathcal{M}})]
\end{equation}
with the operator $\mathcal{D}_1$ being defined as
\begin{align}
    &\mathcal{D}_1[P_t(\mathbf{\mathcal{M}})] := k_m[P_t(\mathcal{M}-\overrightarrow{m}_1)-P_t(\mathcal{M})]  \\
    &+ \sum_{i} \gamma_i [(m_i+1)P_t(\mathcal{M}+\overrightarrow{m}_i)-m_iP_t(\mathcal{M})] \notag \\
    &+ \sum_{i} \sum_{i \neq j} \alpha_{ij}[(m_j+1)P_t(\mathcal{M}-\overrightarrow{m}_i+\overrightarrow{m}_j)-m_jP_t(\mathcal{M})]. \notag
\end{align}
The first and second terms describe mRNA production and degradation{,} respectively{,} while the third term accounts for the mRNA inter-state transitions. Applying the PPA mapping, we rescale $k_m \rightarrow k_m/N$ in the master equation. Since each partition is {constrained to have} either 0 or 1 mRNA when $N\rightarrow \infty$, for the reduced model, let us define a new mRNA state $0$ to represent the absence of the mRNA ({Figure} 1C) and write the corresponding realization of $\mathcal{M}$ as $\overrightarrow{m}_0 = (0,0,...,0)$. Transitions into the $0$ state are simply given by the degradation rates of the mRNA. Transitions out of the $0$ state occur at rate $k_m/N$ and may only go into the $1$ state. From here, it becomes clear that the mRNA states in the reduced model resemble promoter states. Decomposing the master equation with respect to the mRNA states, we define the vector $\overrightarrow{p}_t(\mathcal{M}) = (p_t(\overrightarrow{m}_0), p_t(\overrightarrow{m}_1), p_t(\overrightarrow{m}_2),..., p_t(\overrightarrow{m}_n))$ to rewrite the master equation system in matrix form
\begin{equation}
    \partial_t \overrightarrow{p}_t(\mathcal{M}) = \bigg(\frac{k_m}{N}\boldsymbol{R}+\boldsymbol{A}\bigg)  \overrightarrow{p}_t(\mathcal{M})
\end{equation}
where $\frac{k_m}{N}\boldsymbol{R}+\boldsymbol{A}$ constitutes the inter-state transition rate matrix for the reduced model (of size $n+1$) with $\boldsymbol{R}_{00} = -1$, $\boldsymbol{R}_{10} = 1$, and 0 for all other elements. The block matrix $\boldsymbol{A}$ is given by 
\begin{equation}
    \boldsymbol{A} = 
    \begin{pmatrix}
        0 & \boldsymbol{\Gamma} \\ 
        0 & \boldsymbol{Q} - \text{diag} (\gamma_1, \gamma_2, ..., \gamma_n)
    \end{pmatrix}
\end{equation}
where $\boldsymbol{\Gamma} = (\gamma_1, \gamma_2, ..., \gamma_n)$ and the $n \times n$ matrix

{\begin{equation}
    \boldsymbol{Q} = 
    \begin{cases}
        \alpha_{ij} \,\,\,\,\,\,\,\,\,\,\,\,\,\,\,\,\,\,\,\,\,\,\,\, i \neq j \\
        - \sum_{i \neq j} \alpha_{ij} \,\,\, i = j 
    \end{cases}.
\end{equation}}
We have now reformulated the master equation for the mRNA as a system of equations {that describes} the evolution of a {continuous-time} Markov chain. To find the steady-state distribution, which we denote {by} $^*$, we seek to solve for the nullspace vector by setting
\begin{equation}
    \bigg(\frac{k_m}{N}\boldsymbol{R}+\boldsymbol{A}\bigg) \overrightarrow{p}^*_t(\mathcal{M}) = 0.
\end{equation}
The first row of the matrix equation gives the relation
\begin{equation}
    \frac{k_m}{N}p^*_t(\overrightarrow{m}_0) = \sum_{j=1}^n \gamma_j p^*_t(\overrightarrow{m}_j)
\end{equation} 
and it is helpful to define the weighted average steady-state degradation rate 
\begin{equation}
    \Bar{\gamma} := \frac{\sum_{j=1}^n \gamma_j p^*_t(\overrightarrow{m}_j)}{\sum_{j=1}^n p^*_t(\overrightarrow{m}_j)} = \sum_{j=1}^n \gamma_j V_j
\end{equation}
where $V_j$ represents the probability of finding the {transcript} in state $j$, conditioned on $j \neq 0$,  in the {steady state}.
{
$V_j$ can be obtained from the occupancy of a single transcript before degradation} as the $j$-th element of the vector $\overrightarrow{V}$ that satisfies the matrix equation
\begin{equation}
    (\bold{\Gamma}\bold{I} - \boldsymbol{Q}) \overrightarrow{V} = (\Bar{\gamma},0,0,...)^T
\end{equation}
with $\bold{I}$ being an identity matrix. In conjunction with the normalization condition that $p^*_t(\overrightarrow{m}_0)+\sum_{j=1}^n p^*_t(\overrightarrow{m}_j) = 1$, we now obtain
\begin{align}
    p^*_t(\overrightarrow{m}_0) &= \frac{ \Bar{\gamma} N }{ \Bar{\gamma} N + k_m },   \\
    p^*_t(\overrightarrow{m}_j) &= \frac{k_m V_j}{\Bar{\gamma} N + k_m} .
\end{align}
With $p^*_t(\mathcal{M})$ in hand for all realizations of $\mathcal{M}$, we can simply construct the stationary generating function for the reduced model using its definition
\begin{align}
    g^*(z_1, z_2, \dots, z_n) &= \frac{\Bar{\gamma} N}{\Bar{\gamma}N+k_m}  + \frac{k_m}{\Bar{\gamma} N + k_m} \sum_{j=1}^n z_j V_j
\end{align}
where $z_j$ corresponds to the random variable {representing} the count of $j$-type mRNA. The corresponding generating function for the original model via Eq. (4) is then
\begin{equation}
    G^*(z_1, z_2, \dots, z_n) = \exp \bigg\{ \frac{k_m}{\Bar{\gamma}} \bigg( \sum_{j=1}^n z_j V_j -1 \bigg) \bigg\}
\end{equation}
which is its most general form for $n$ arbitrary post-transcriptional regulatory states. In the limiting case where no post-transcriptional regulatory mechanisms are present where there is only one mRNA state, we have that $\sum_{j=1}^1 z_j V_j = z_1${,} causing Eq. (21) to reduce to the generating function of a Poisson distribution with mean $k_m/\gamma_1$
\begin{equation}
    G^\ast(z_1) = \exp \bigg\{ \frac{k_m}{\gamma_1} (z_1-1) \bigg\}
\end{equation} 
where $\gamma_1$ is the unregulated mRNA degradation rate, thereby recovering a trivial result. Additionally, Eq. (21) is also consistent {with queueing theory, where} the number of mRNA in the system maps onto the number of customers waiting in a queue \cite{Jia-PRL-2011}. Given that mRNA lifetimes are {independent, identically distributed} random variables, the model under consideration can be taken to be what is known as the $M/G/\infty$ queue {using Kendall's notation} in the queueing theory literature. {The $M$ represents that arrivals follow a Markovian process (i.e. exponentially distributed wait-times), $G$ indicates that the wait-time between departures can take on a general distribution, and $\infty$ indicates an infinite server queue to reflect that degradation events occur independently of on another.} For a comprehensive discussion on the connection between gene expression models and queueing systems, see Ref. \cite{Nossan-BJ-2024}. A classical result for the $M/G/\infty$ queue is that its steady-state queue length follows a Poisson distribution \cite{Newell-SIAM-1966}. As such, when we consider the \textit{marginal} mRNA count (i.e. total mRNA count without distinguishing between different mRNA states) by taking the limit $z_j \rightarrow z$ for all $j$ and noting that $\sum_{j} V_j = 1$, we see that Eq. (21) similarly reduces to
\begin{align}
    G^\ast(z) &= \lim_{z_1 \rightarrow z} \dots \lim_{z_n \rightarrow z} 
    G^\ast(z_1, z_2, \dots, z_n) 
    \notag \\
    &= \exp \bigg\{ \frac{k_m}{\bar{\gamma}} (z-1) \bigg\}
\end{align}
which is the generating function of the Poisson distribution with mean $k_m/\bar{\gamma}$ as expected.

\subsection{Exact Results for Protein Distributions}
Adding a layer of complexity, let $H$ be the random variable representing the number of proteins. We define the mRNA-protein joint distribution $P_t(\mathcal{M},H)$ whose master equation is given by 
\begin{align}
    \partial_t P_t(\mathbf{\mathcal{M}},H) = \mathcal{D}_1[P_t(\mathbf{\mathcal{M}},H)] +\mathcal{D}_2[P_t(\mathbf{\mathcal{M}},H)]. 
\end{align}
{Here,} $\mathcal{D}_1$ was previously given by Eq. (10) and the operator $\mathcal{D}_2$ is
\begin{align}
   \mathcal{D}_2[&P_t(\mathbf{\mathcal{M}},H)] := \sum_{i} \kappa_i[m_iP_t(\mathcal{M},H-1)-m_iP_t(\mathcal{M},H)] \notag \\
   &+\mu_p[(H+1)P_t(\mathcal{M},H+1) -{H}P_t(\mathcal{M},H)].
\end{align}
The two terms in Eq. (25) account for protein production and degradation in their respective order. Once again, after applying the PPA mapping, $k_m \rightarrow k_m/N$ and enforcing $N \rightarrow \infty$, each partition is constrained to have either 0 or 1 mRNA{,} so we can treat the reduced model for protein production as a promoter-based model for mRNA production studied in Ref. \cite{Sanchez-PNAS-2008}. Specifically, the protein production rates for each mRNA state in our model map onto the mRNA production rates in the {promoter-based} model. The degraded mRNA (state 0) ``promoter'' state will have a protein production rate of $0$. 

As originally noted \cite{Pendar-PRE-2013}, the significance of the PPA mapping here lies in the fact that any result for the mRNA distribution for a promoter-based model directly translates to a result for the protein distribution of a corresponding model of post-transcriptional regulation. The reduced models that arise here are MAPs whose protein levels map onto the $G/M/\infty$ queue {(general arrival process, Markovian degradation process, and infinite server queue)}, from which $g^\ast(z)$ can be obtained. Prior work on the $G/M/\infty$ queue has shown that if the wait-time distribution between arrival events can be analytically obtained in the Laplace domain as a rational function, then an exact stationary generating function follows \cite{Nossan-BJ-2024}{,} which directly translates to an exact result for $G^\ast(z)$. Nonetheless, obtaining the exact moments of the protein distribution is more flexible and straightforward to use when we are solely interested in quantifying noise.

{Turning to} the exact expressions for the moments of the protein distribution, we decompose Eq. (24) into a system of equations with respect to the mRNA states, similar to the example presented in the previous subsection. It has been shown that the vector $\overrightarrow{x}^{(j)}$ that encodes the arbitrary $j$-th steady-state moments of the protein distribution for each mRNA state can be found {by} iteratively solving \cite{Sanchez-PNAS-2008} 
\begin{equation}
     \bigg(j\mu_p \bold{I} - \bigg(\frac{k_m}{N} \bold{R}+\bold{A}\bigg)\bigg) \overrightarrow{x}^{(j)} = j \bold{K} \overrightarrow{x}^{(j-1)}.
\end{equation}
{Here,} the diagonal elements of $\bold{K}$ encode the protein production rates of each mRNA state. Since we are interested in the marginal protein statistics, summing over all possible realizations of $\mathcal{M}$ leads to the iterative relation for the factorial moments of $P^*(H)$
\begin{equation}
     \bigg\langle \frac{H!}{(H-j)!} \bigg\rangle = \overrightarrow{1} \cdot \overrightarrow{x}^{(j)} = \frac{\overrightarrow{\kappa} \cdot \overrightarrow{x}^{(j-1)}}{\mu_p}
\end{equation}
where $\overrightarrow{\kappa} = (0, \kappa_1, \kappa_2, ..., \kappa_n)$. The initial $\overrightarrow{x}^{(0)}$ needed to compute the {higher-order} moments can be found by solving $(\frac{k_m} {N}\boldsymbol{R}+\boldsymbol{A})\cdot\overrightarrow{x}^{(0)} = 0${,} whose solution was discussed in the previous subsection. These results in conjunction with the established relationship between the moments of the reduced and original model {constitute} an exact, analytical framework for analyzing post-transcriptional regulation in \textit{general} models.

\section{Transcriptional Bursting}
At the transcriptional level, mRNA molecules are often produced sporadically in batches \cite{Kumar-PlosCompBio-2015,Elgart-PhysBiol-2011}. This is referred to as transcriptional bursting and has been observed across a broad class of genes \cite{Golding-Cell-2005,Suter-Science-2011}. Regulation of transcriptional bursting involves modifying the mean burst size and/or its arrival process. In this case, {unlike before},  mRNA arrivals now follow a \textit{batch} Poisson arrival process with the burst size being drawn from the distribution $P(b)${,} where $b$ is the random variable corresponding to the burst size. The post-transcriptional regulatory mechanisms that affect the mRNA remain arbitrary and are likewise depicted {in} Figure 1A. Thus, the lifetimes of mRNA produced between different bursts remain {independent, identically distributed} random variables.

Let $P(h,b)$ be the joint distribution in the reduced model for the protein count $h$ generated from an mRNA burst of size $b$. We note that in a single burst, the $b$ mRNAs enter into the reduced model simultaneously. The key is that we only need to consider the arrival of one burst for the reduced model since the probability of more than one burst arrival is negligible by the nature of the PPA mapping. Although protein arrivals still follow the $G/M/\infty$ queue, it is unclear how the waiting time distribution between protein arrivals can be obtained given that $b$ is a random variable, so the queueing theory approach from Ref. \cite{Nossan-BJ-2024} {cannot be used to derive} the exact protein distribution. However, the first two moments are still analytically tractable. Now, for each partition{,} we condition on the burst size $b$
\begin{equation}
    P(h,b) = P(b)P(h\vert b) 
\end{equation}
with the corresponding generating function of $P(h,b)$, 
\begin{equation}
    G(z) = \sum_{b=0}^\infty P(b) G_b(z)
\end{equation}
where we denote $G_b(z) := \sum_{h=0}^\infty z^h P(h\vert b)$ {as} the generating function for the partition conditional on {burst size} $b$. Eq. (29) can be interpreted as the weighted average of $G_b(z)$ with respect to the likelihood of $b$. For the first moment of $h$, we have
\begin{equation}
    \langle h \rangle = \partial_z G(z) \vert_{z=1} =  \sum_{b=0}^\infty P(b) \langle h_b \rangle
\end{equation}
where $\langle h_b \rangle$ is the expected value of the number of proteins produced by $b$ mRNAs simultaneously. Since $h_b$ is the sum of $b$ identically distributed random variables, $\langle h_b \rangle = b \langle h_1 \rangle$ where $h_1$ is the random variable corresponding to the number of proteins produced when there is {one mRNA molecule}. Substituting this into Eq. (30), the first moment comes out to
\begin{equation}
    \langle h \rangle = \langle h_1 \rangle \sum_{b=0}^\infty b \cdot P(b) = \langle h_1 \rangle \langle b \rangle{.}
\end{equation} 

{Continuing with} the second moment, let us start with the identity
\begin{equation}
    \langle h(h-1) \rangle = \partial_z^2 G(z)\vert_{z=1} = \sum_{b=0}^\infty P(b) \langle h_b(h_b-1) \rangle.
\end{equation}
For the reduced model, {in the limit} $N \rightarrow \infty${, where} $\langle h \rangle \propto 1/N ${,} $N\langle h \rangle^2$ {vanishes; thus,} the variance of $h$ can be simply computed as $\langle h^2 \rangle = (\partial^2_z G(z)+ \partial_z G(z)) \vert_{z=1}$. This also applies to the random variables $h_b$ for all $b$, which leads to
\begin{equation}
    \text{Var}\{h\} = \sum_{b=0}^\infty P(b) \langle h_b^2\rangle
\end{equation}
where we have combined equations (30) and (32). Since {mRNA molecules} from a single burst arrive simultaneously, the random variables corresponding to the number of proteins produced from each mRNA strand will be correlated{,} so the variance for $h_b$ is
\begin{equation}
    \langle h_b^2\rangle = b \langle h_1^2 \rangle + b(b-1) \text{Cov} \{ h^{(i)}, h^{(j)} \}
\end{equation}
where $\{ h^{(i)}, h^{(j)} \}$ can be taken to be any pairwise combination of the identically distributed random variables ($i$ and $j$ give the index of an individual mRNA molecule). Note that $b(b-1)$ gives the number of possible pairwise combinations of {mRNA molecules}. Since the covariance term {is} identical for all mRNA pairs{ because} $\{ h^{(i)}, h^{(j)} \}$ {are} exchangeable random variables for all $\{i,j \}$ given $i\neq j$, we can make use of the $b=2$ case to determine the covariance term 
\begin{equation}
    \langle h_2^2 \rangle = 2\big( \langle h_1^2 \rangle + \text{Cov}\{ h^{(1)}, h^{(2)} \}\big)
\end{equation}
which obviously rearranges to
\begin{equation}
    \text{Cov}\{ h^{(1)}, h^{(2)} \} = \frac{\langle h_2^2 \rangle}{2} - \langle h_1^2 \rangle.
\end{equation}
Substituting Eq. (36) into Eq. (34) and then the resulting expression into Eq. (33), the second moment works out to be
\begin{align}
    \text{Var}\{ h \} = \langle b(b-1) \rangle \bigg[ \frac{\langle h^2_2 \rangle}{2} - \langle h_1^2 \rangle \bigg] + \langle b \rangle \langle h_1^2 \rangle.
\end{align}
{Using the PPA relation between reduced and original moments, the protein Fano factor is given by}
\begin{equation}
    F = F_1 + \frac{\langle b(b-1) \rangle}{\langle b \rangle} \big( F_2 - F_1 \big).
\end{equation}
Remarkably, the Fano factor at the protein level for the full model can be completely determined by just finding the Fano factors for the simple cases {in which} exactly {one} ($F_1$) and {two} ($F_2$) mRNA molecules simultaneously enter into the system (which we will refer to as the $b=1$ and $b=2$ burst cases{,} respectively) for any underlying transcriptional burst distribution. The $b=1$ case corresponds to simple Poisson mRNA arrivals{, and we consider} a separate kinetic scheme for the reduced model for the $b=2$ burst case{,} which we will {illustrate} in the following section.

\section{Applications}
In this section, we demonstrate the versatility of the PPA mapping by deriving analytical results for a diverse set of biologically relevant models. {The recipe for applying the PPA mapping on a model of post-transcriptional regulation to obtain the Fano factor is as follows and tracks closely with the framework from Ref. \cite{Sanchez-PNAS-2008}:
\begin{enumerate}
    \item Construct the reduced model by adding $m_0$ to the set of mRNA states and set the transition rate from $m_0$ to a free mRNA state be $k_m/N$ and route all the mRNA degradation pathways back to $m_0$. The transition dynamics of the singular mRNA here is governed by the matrix $\frac{k_m}{N}\boldsymbol{R}+\boldsymbol{A}$. 
    \item Solve for $(\frac{k_m}{N}\boldsymbol{R}+\boldsymbol{A})\overrightarrow{x}^{(0)} = 0$. 
    \item Using the solution for $\overrightarrow{x}^{(0)}$, solve: \begin{equation}
         \bigg(\mu_p \bold{I} - \bigg(\frac{k_m}{N} \bold{R}+\bold{A}\bigg)\bigg) \overrightarrow{x}^{(1)} =  \bold{K} \overrightarrow{x}^{(0)}. \notag 
    \end{equation}
    \item Calculate the factorial moments: $\langle h \rangle =  \overrightarrow{1} \cdot \overrightarrow{x}^{(0)}$ and $ \langle h(h-1) \rangle =  \overrightarrow{1} \cdot \overrightarrow{x}^{(1)}$.
    The Fano factor for a single transcript follows as: $F_1 = 1 + \langle h(h-1) \rangle /\langle h \rangle$.
    \item If there is transcriptional bursting, obtain $F_2$ by repeating steps 2-4 on the kinetic scheme for the $b=2$ arrival case and determine $\langle b(b-1) \rangle$ and $\langle b\rangle$ for the given burst distribution. The Fano factor now follows from Eq. (38). We give an illustration of handling the $b=2$ case in Section V.B. 
\end{enumerate}} Each example chosen demonstrates a different capability of the PPA mapping. 

\subsection{Multi-step mRNA Degradation}
mRNA degradation is often assumed to be a one-step, first-order process corresponding to exponentially distributed lifetimes. However, realistically, mRNA degradation is a multi-step process {in which} separate parts of the mRNA molecule are {degraded successively}. The majority of mRNA species' lifetime distributions are empirically shown to deviate from {an exponential distribution} \cite{Deneke-PLOS-2013}. In the model under consideration (left panel of Figure 2){,} after an mRNA molecule is produced, it undergoes a successive $n$-step degradation sequence, where the protein production rate for each step is $k_p$ and transition rates {between} them are $\mu_m$. mRNA molecules are produced at a rate of $k_m${;} proteins degrade at rate $\mu_p$, and we take $n$ to be arbitrary. In this case, mRNA lifetimes follow a gamma distribution with mean $\frac{n}{\mu_m}$. 

\begin{figure*}
    \centering
    \includegraphics[width=\linewidth]{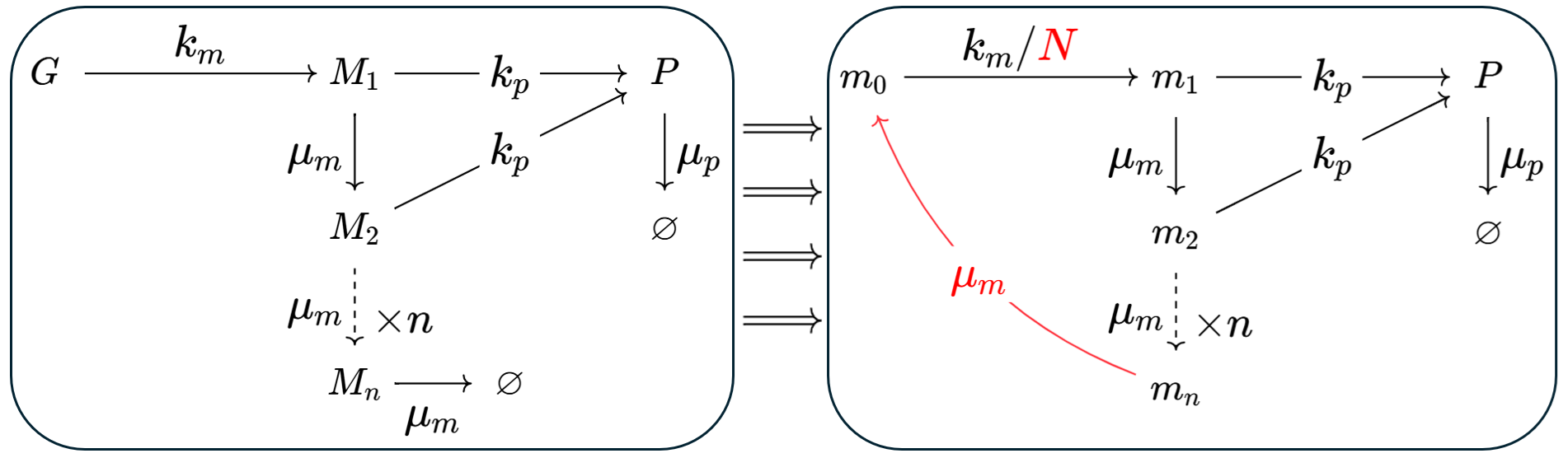}
    \caption{Schematic of mRNA undergoing arbitrary $n$-step degradation mapped onto its reduced model. {mRNA molecules} are produced at rate $k_m$ and transition rates {between degradation steps} are $\mu_m$. Each degradation step produces proteins at rate $k_p$, and proteins degrade at rate $\mu_p$.}
\end{figure*}

Results for this model were previously derived in Ref. \cite{Pedraza-Science-2008}{, and,} following the PPA framework detailed in this paper, we construct its reduced model (right panel of Figure 2) where the {mean and} Fano factor at the protein level works out to be:
\begin{equation}
    {\langle H \rangle = \frac{nk_mk_p}{\mu_m\mu_p}} \notag 
\end{equation}
\begin{align}
    F = 1 + \frac{k_p}{\mu_p} \bigg[ 1 - \frac{\mu_m}{\mu_p}\frac{1}{n}  \bigg[ 1 - \bigg( \frac{\mu_m}{\mu_p+\mu_m} \bigg)^n \bigg] \bigg]
\end{align}
which is in agreement{,} as expected{. The} full derivation is in Appendix A. Remarkably, the size of the model, taken to be arbitrary, can be analytically incorporated as a parameter in the PPA framework.

\subsection{Synthetic Noise Control using sRNAs}
For the design of effective gene circuits in synthetic biology, it is crucial to minimize noise in the mRNA and protein counts to prevent crosstalk, which is the tendency for signals from one gene circuit {to} unintentionally interact and interfere with another. Buffering noise is critical for systems where the underlying genes exhibit transcriptional bursting. sRNAs, a class of non-coding RNAs, have emerged as a powerful tool for synthetically controlling gene expression \cite{Mehta-MSB-2008}. sRNAs modulate gene expression by binding to an mRNA, forming a complex with a modified translation and/or degradation rate. 

\begin{figure*}
    \centering
    \includegraphics[width=\linewidth]{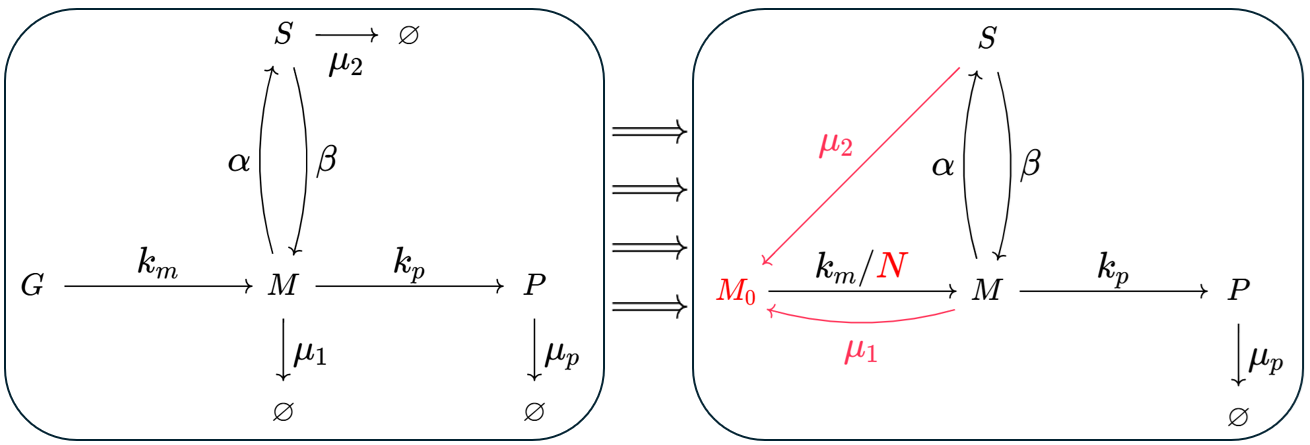}
    \caption{Kinetic scheme of sRNA regulation mapped onto its reduced model. G: gene; M: unbound mRNA; S: sRNA-mRNA complex; P: protein; M$_0$: ``degraded{''} promoter state for {the} reduced model.} 
\end{figure*}
\begin{figure}
    \centering
    \includegraphics[width=\linewidth]{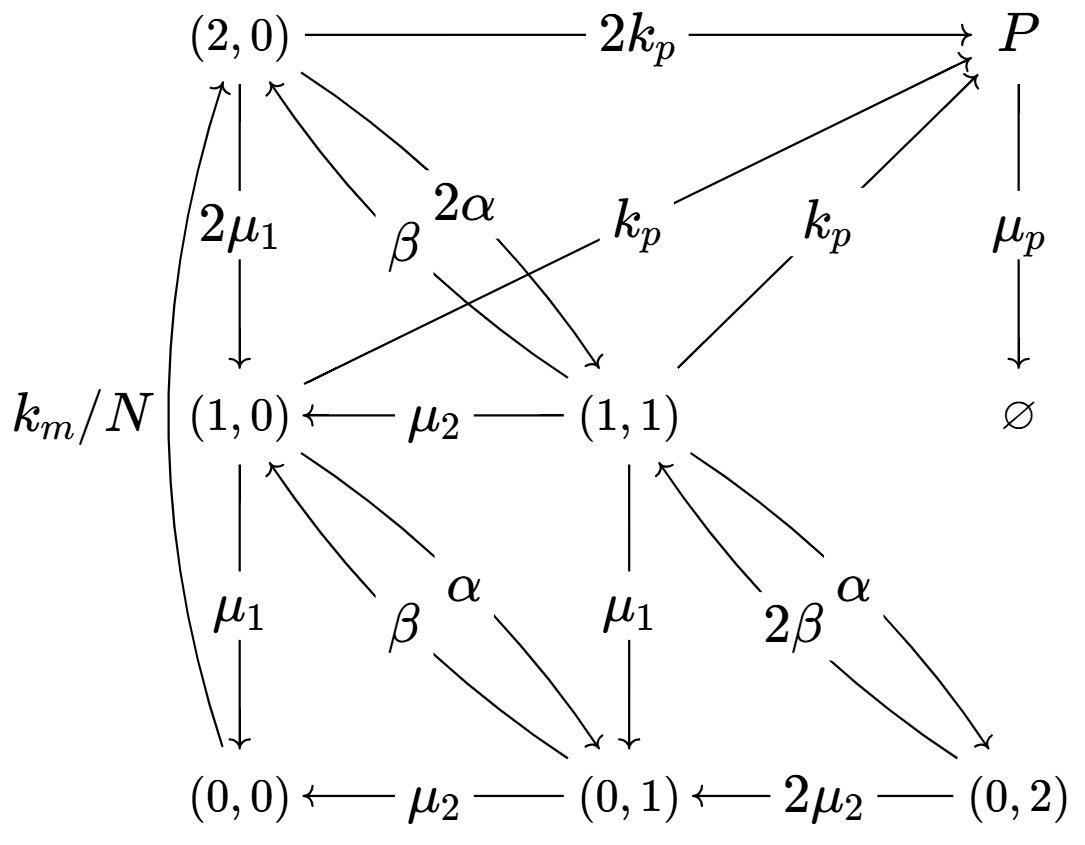}
    \caption{Kinetic scheme to be considered for the $b=2$ case {in which two mRNA molecules} simultaneously enter into the system. The states are specified by the ordered pair $(f,c)$ where $f$ is the number of free mRNA molecule(s) and $c$ represents the number of sRNA-mRNA complex(es). The protein production rate from each state is $k_pf$.}
    \label{figplaceholder}
\end{figure}

Consider the minimal model of sRNA regulation depicted in {the} left panel of Figure 3. mRNA arrive in transcriptional bursts at rate $k_m$ with the burst size drawn from the geometric distribution \cite{Elgart-PhysBiol-2011}{,} which takes the generating function form
\begin{equation}
    G_b(z) = \frac{1}{1+\langle b\rangle (1-z)}
\end{equation}
where $b$ is the random variable corresponding to burst size. A single mRNA molecule binds to {an} sRNA molecule at rate $\alpha$, forming an mRNA-sRNA complex {that} fully represses the mRNA (i.e. the complex cannot translate). The unbinding rate of the complex is $\beta$. We assume that individual mRNA-sRNA binding and unbinding events have a negligible effect on the overall sRNA concentration, allowing us to take $\alpha$ to be constant. The unbound mRNA and mRNA-sRNA complex degrade at rate $\mu_1$ and $\mu_2${,} respectively. Unbound mRNA translate at rate $k_p$ and proteins degrade at rate $\mu_p$. 

Following the PPA mapping detailed in the previous sections, the model maps onto its reduced counterpart (right panel of Figure 3) and we arrive at the following expression for the steady-state {mean and} Fano factor of the protein distribution 
\begin{widetext}
    \begin{equation}
            {\langle H \rangle = \frac{\langle b \rangle k_mk_p(\beta+\mu_2)}{\mu_p(\beta \mu_1 + \mu_2(\alpha+\mu_1))}} \notag 
    \end{equation}
    \begin{align}
        F = 1+ \frac{{k_p} \big[\alpha  {\mu_p} (\langle b \rangle {\mu_2}+\beta +{\mu_2})+\alpha (\langle b \rangle +1) (\beta +{\mu_2})^2 +(\langle b \rangle +1) (\beta +{\mu_2}) (\beta +{\mu_1}+{\mu_2}) (\beta +{\mu_2}+{\mu_p})\big]} {\big(({\mu_2}+{\mu_p}) (\alpha +{\mu_1}+{\mu_p})+\beta  ({\mu_1}+{\mu_p}) \big) (\beta +{\mu_2}) (\alpha +\beta +{\mu_1}+{\mu_2})}
    \end{align}
\end{widetext}
The full details of the calculation are given in Appendix B. To give an overview, we needed to derive the first and second factorial moments for both the $b=1$ and $b=2$ cases. As noted before, the $b=1$ case corresponds to mRNA arrivals following a simple Poisson process{,} and the operations performed on the transition matrix of the reduced model were exactly the same as {in} the previous example. For the $b=2$ case, we had to consider the kinetic scheme depicted in Figure 4{, in which two} mRNA molecules simultaneously enter into the system. Each state is denoted by an ordered pair $(f,c)${, where} $f$ and $c$ represent the {numbers} of free and bound {mRNA molecules} associated with the state{,} respectively{,} and the corresponding protein production rate is $k_pf$. The calculations carried out with the kinetic scheme depicted in Figure 4 are identical to what we have done with the $b=1$ case. Note that similar kinetic schemes can {also} be constructed to analyze greater burst sizes using the same logic. In the limiting case where $\alpha \rightarrow 0$, the model collapses to the well-known two-stage model of gene expression with geometrically distributed transcriptional bursts and with the Fano factor
\begin{equation}
    \lim_{\alpha \rightarrow 0 } F = 1 + \frac{k_p(1+\langle b \rangle)}{\mu_1+\mu_p},
\end{equation}
recovering a trivial result \cite{Shahrezaei-PNAS-2008, Pedraza-Science-2008}. {For another consistency check, in the $\beta \to 0$ limit where sRNA binding is irreversible, once an mRNA is bound it may never translate again so the model similarly reduces to a two-stage model with geometrically distributed transcriptional bursts with an effective mRNA degradation rate of $\mu_1+\alpha$. Correspondingly, 
\begin{equation}
    \lim_{\beta\to 0} F = 1 + \frac{k_p(1+\langle b\rangle)}{\mu_1+\alpha+\mu_p}
\end{equation}
which differs by a term of $\alpha$ in the denominator from Eq. (42) as expected.}

\begin{figure}
    \centering
    \includegraphics[width=1\linewidth]{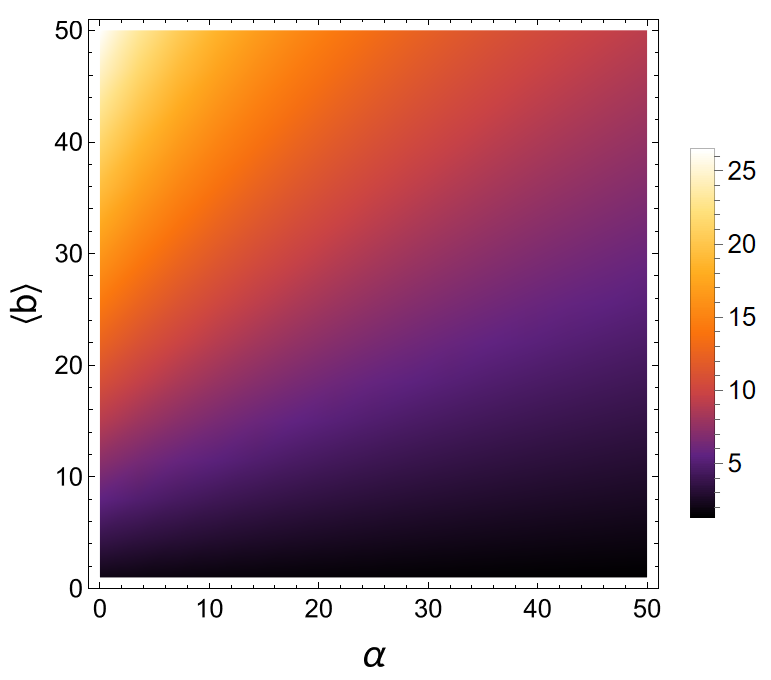}
    \caption{{Density plot of protein noise regulation with sRNA, showing} the Fano factor Eq. (41) as a function of $\alpha$ and $\langle b\rangle$. Parameters $k_m = k_p = \mu_1 = \mu_p = \beta = 10 \,s^{-1}$; $\mu_2 = 40\,s^{-1}$. }
    \label{figplaceholder}
\end{figure}

Plotting out the Fano factor (Figure 5) as a function of $\alpha$ and $\langle b \rangle$, we readily see that increasing the average burst size increases the noise, but can be correspondingly buffered by increasing the sRNA expression level and vice versa; the two mechanisms can be used in conjunction to achieve the desired quantitative protein noise level. {This can lead to synthetic control strategies that utilize precise control of transcriptional burst sizes and sRNA expression levels for minimizing fluctuations in gene circuit output to prevent unwanted effects such as crosstalk \cite{Bucc-NatGeneRev-2020}.} Remarkably, when $\alpha$ is large compared to $\langle b \rangle$, the variability in the protein counts is nearly Poissonian. In fact, taking the limit $\frac{\alpha}{\langle b \rangle} \rightarrow \infty$ gives $F=1$.

\subsection{Exact Protein Distribution with Nuclear Export of mRNA} 
Nuclear retention mechanisms serve as quality control for mRNA and buffer gene expression in rapidly changing environmental conditions \cite{Halpern-CellRep-2015, Smith-IEEE-2021}. Before export to the cytoplasm, mRNA in the nucleus must undergo a set of preprocessing steps such as the splicing of the introns, 5' capping, 3' polyadenylation, etc., resulting in a stochastic delay which affects the variability in mRNA and protein levels. Let us assume that the wait-time distribution between export events is exponentially distributed \cite{Singh-BioJ-2012}. Consider the kinetic scheme of nuclear retention depicted {in} Figure 6. mRNA molecules are produced at a constant rate $k_m$. After being produced, the mRNA initially reside in the nucleus and must be exported to the cytoplasm, which occurs at rate $k_E$, before translation can take place. mRNA in the cytoplasm degrade at rate $\mu_m$ and produce proteins at rate $k_p$. Proteins degrade at rate $\mu_p$. We will proceed to use the PPA mapping to derive the stationary protein distribution. Though we give a brief summary of the derivation here, the full mathematical details are given in Appendix C. 

\begin{figure}
    \centering
    \includegraphics[width=0.9\linewidth]{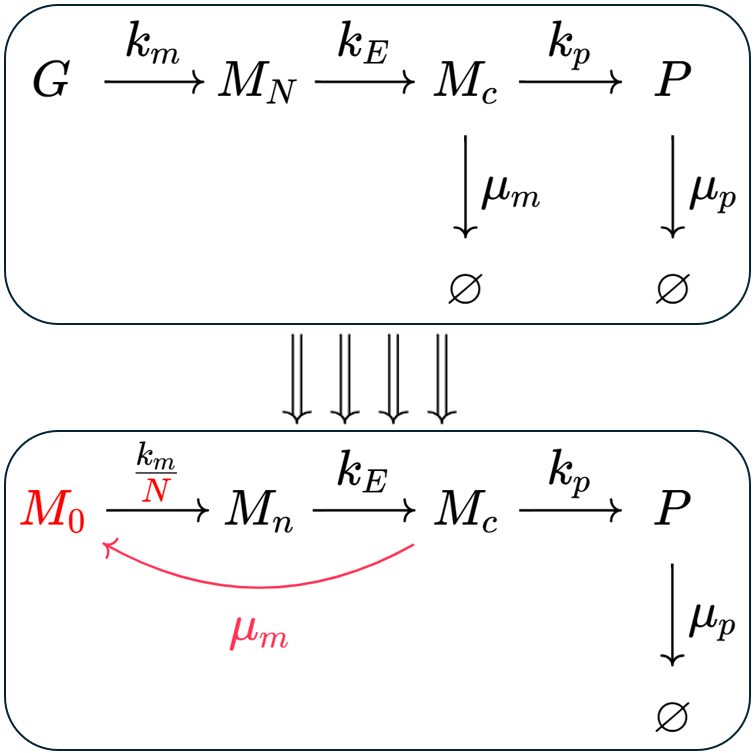}
    \caption{Schematic of protein production {in which} the mRNA are subject to nuclear export. $G$ gene; $M_n$ nuclear mRNA, $M_c$ cytoplasmic mRNA; $P$ protein.}
    \label{figplaceholder}
\end{figure}

From Figure 6, we see that the reduced model resembles a 3-state refractory promoter model {in which} only one-way transitions occur between adjacent states. As previously noted, protein production and degradation map onto the $G/M/\infty$ queue and the exact stationary distribution can be obtained using a queueing theory approach. Let $\phi(s)$ denote the Laplace transform of the interarrival time distribution between protein production events{,} which we derive to be
\begin{align}
    &\phi(s) = \frac{k_p(k_E+s)( \frac{k_m}{N}+s)}{(k_E+s)(\frac{k_m}{N}+s)(k_p+s) + s\mu_m(k_E+\frac{k_m}{N}+s)}.
\end{align}
The stationary generating function of a $G/M/\infty$ queue corresponding to a MAP can be neatly expressed as a generalized hypergeometric function{. Following} Ref. \cite{Nossan-BJ-2024}, the parameters are obtained {by} factoring the rational expression $\phi(s)/(1-\phi(s))$. In conjunction with Eq. (4), this leads to the stationary generating function for the protein distribution of the original model
\begin{align}
    G^\ast(z) = \lim_{N \rightarrow \infty} &\exp \bigg\{ N \bigg( {_2F_2}\bigg[ \frac{k_E}{\mu_p}, \frac{k_m/N}{\mu_p}; \notag \\ 
    &\frac{b_1}{2\mu_p}, \frac{b_2}{2\mu_p}; \frac{k_p}{\mu_p}(z-1) \bigg] - 1 \bigg) \bigg\},
\end{align} 
where
\begin{align}
    b_1 &= k_E + \frac{k_m}{N} + \mu_m + \sqrt{\Delta}; \\
    b_2 &= k_E + \frac{k_m}{N} + \mu_m - \sqrt{\Delta}; \\
    \Delta &= \bigg( k_E - \frac{k_m}{N} \bigg)^2 - 2\mu_m\bigg( k_E + \frac{k_m}{N} \bigg) + \mu_m^2.
\end{align}
{We benchmarked Eq. (45) against stochastic simulations using Gillespie's algorithm \cite{Gillespie-JCompPhys-1976} for an intermediate parameter regime away from any limiting case, which shows near-perfect agreement (Figure 7). The exact probability distribution plotted is obtained via the Taylor series of its generating function counterpart $\frac{1}{n!} [ \partial^nG(z)/\partial z^n ]\vert_{z=0}$.}
\begin{figure}
    \centering
    \includegraphics[width=1\linewidth]{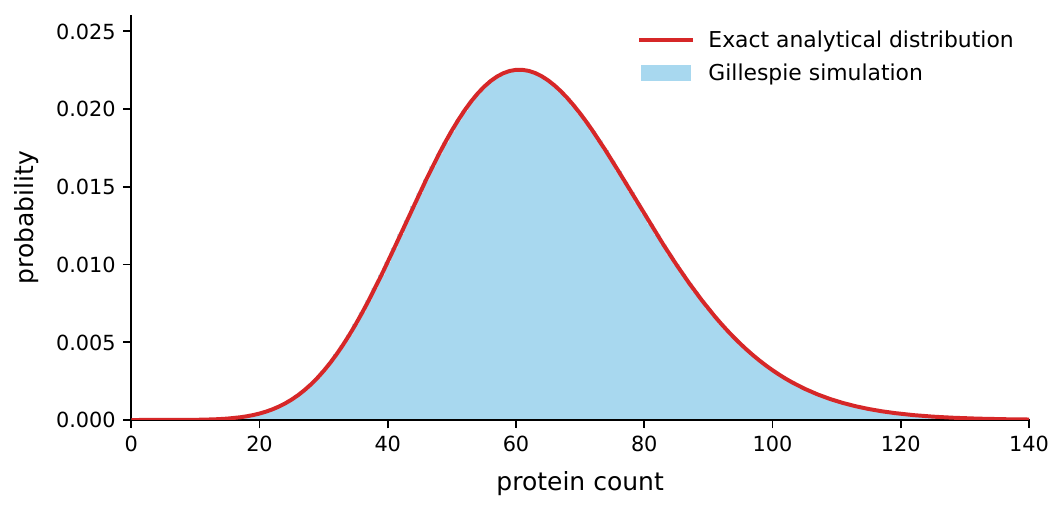}
    \caption{{Steady-state protein distribution for the nuclear export model depicted in Figure 6 using the parameters: $k_m=k_p=80$\,$s^{-1}$ and $\mu_m=\mu_p=k_E=10$\,$s^{-1}$.}}
\end{figure}

\section{Discussion}
Experimental efforts to study variability in gene expression have  increasingly emphasized regulation beyond promoter switching, including post-transcriptional mechanisms such as sRNA regulation, nuclear retention, and processing-dependent transcript lifetimes \cite{Hansen-BioE-2019, Cai-GPB-2009, Halpern-CellRep-2015, Furlan-BinB-2021}. These mechanisms are much more structurally diverse than standard promoter-state models, which makes it useful to have a common analytical reduction. We have shown that the PPA mapping provides such a reduction for finite-state post-transcriptional models with linear reaction rates{. After} partitioning the transcript arrivals and taking $N\to\infty$, each partition is {constrained to contain} either zero or one transcript, causing the transcript's post-transcriptional states {to become} the states of an auxiliary promoter model.

This mapping has two practical consequences.  First, the steady-state protein factorial moments can be obtained by solving a fixed $(n+1)$-state linear system recursively, rather than by writing a set of increasingly large coupled moment equations \cite{Singh-IEEE-2007}. Second, when the reduced protein-arrival process has a rational Laplace transform, recent results for the $G/M/\infty$ queue can be used to obtain the full stationary protein generating function \cite{Weidemann-SciAdv-2023, Muhan-MathBio-2024, Wang-PRL-2025, Nossan-BJ-2024}. Importantly, the method does not require the fast-mRNA approximation in which each transcript is replaced by an instantaneous burst of proteins, complementing our earlier coarse-grained analyses of post-transcriptional noise control \cite{Jia-PRL-2010, Baker-PRE-2012, Kumar-PlosCompBio-2015, Kumar-PhysBiol-2019}. The models chosen in the Applications section above illustrate these points for multi-step mRNA degradation, sRNA-mediated buffering of bursting-induced noise, and nuclear export. In synthetic gene circuits, such formulas can be useful for tuning mean expression and noise {to reduce} unwanted crosstalk \cite{Swan-Nature-2011, Bucc-NatGeneRev-2020}.

Furthermore, the extension to transcriptional bursting shows that burst-size statistics can be incorporated into the PPA framework in a particularly compact way. For a batch Poisson transcription process with burst size distribution $P(b)$, Eq. (38) expresses the protein Fano factor in terms of only two ingredients: the first two factorial moments of $P(b)$ and the two reduced post-transcriptional response calculations corresponding to the $b=1$ and $b=2$ cases. The $b=1$ problem gives the protein noise generated by an isolated transcript, whereas the $b=2$ problem encodes the pairwise covariance between proteins produced by two transcripts born in the same burst. Thus, rather than requiring a separate calculation for each possible burst size, the full burst size distribution enters in Eq. (38) only through $\langle b(b-1)\rangle/\langle b\rangle$, providing a substantial simplification.

In the absence of post-transcriptional regulation, Eq. (38) reduces to the known result for protein noise with Poisson burst arrivals and arbitrary burst-size statistics \cite{Pedraza-Science-2008}. Our result extends this structure by replacing the simple transcript lifetime response with a general post-transcriptional response kernel. In this sense, Eq. (38) provides a controlled generalization of the Poisson-burst-arrival case{:} burst statistics determine how strongly simultaneously produced transcripts contribute to protein noise, while the reduced $b=1$ and $b=2$ problems determine how the post-transcriptional dynamics transduce those bursts into protein fluctuations. {We do note, however, that gene networks are also often extensively regulated at the transcriptional level which can cause mRNA burst arrivals to deviate from a Poisson process \cite{Kumar-CompBio-2015, Pedraza-Science-2008, Jia-PRL-2011}. A limitation of our results is that extending the PPA framework into this setting is unclear.} 

In conclusion, the PPA mapping provides a unifying analytical bridge between post-transcriptional transcript dynamics and promoter-state models, yielding a computationally tractable framework for calculating protein moments and, in certain cases, full stationary protein distributions, in stochastic gene-expression models that combine transcriptional bursting with post-transcriptional regulation.

\section*{Acknowledgments}
We thank Thierry Platini for helpful input during the early stages of the work.

\setcounter{equation}{0}
\renewcommand{\theequation}{A\arabic{equation}}

\onecolumngrid

\section*{Appendix A: Multi-step mRNA Degradation Fano Factor}
Let us first construct the transition rate matrix $T$ of the reduced model
\begin{equation}
    T=
    \begin{bmatrix}
        \frac{-k_m}{N}  & 0 & 0 & 0 & \dots & \mu_m  \\ 
        \frac{k_m}{N}  & -\mu_m & 0 & 0 &  &   \\ 
        0 & \mu_m & -\mu_m & 0 &  &   \\ 
        0 & 0 & \mu_m & -\mu_m &  &  \\ 
        \vdots &  &  & \ddots & \ddots &   \\ 
        0 &  &  &  & \mu_m &  -\mu_m \\ 
    \end{bmatrix}
\end{equation}
whose nullspace vector $\overrightarrow{x}^{(0)}$ encodes the stationary distribution of the mRNA states{,} which we find to be
\begin{equation}
    \overrightarrow{x}^{(0)} = \bigg( \frac{\mu_m}{\mu_m + n (k_m/N)}, \frac{k_m/N}{\mu_m + n (k_m/N)}, \frac{k_m/N}{\mu_m + n (k_m/N)}, \dots, \frac{k_m/N}{\mu_m + n (k_m/N)}\bigg)^\top.
\end{equation}
Taking the inner product described in Eq. (27) from the main text where $\overrightarrow{\kappa}=(0,k_p,k_p,\dots,k_p)$, the mean protein count of the reduced model works out to
\begin{equation}
    \langle h \rangle = \frac{k_p}{\mu_p}\bigg[ \frac{n(k_m/N)}{\mu_m+n(k_m/N)} \bigg].
\end{equation}
To iterate Eq. (27) again for the second factorial moment, we need to find the elements of $\overrightarrow{x}^{(1)}$ beforehand. Following Eq. (26) of the main text where we set $j=1$, we construct the following system of equations
\begin{equation}
    \begin{bmatrix}
        \mu_p+\frac{k_m}{N}  & 0 & 0 & 0 & \dots & -\mu_m  \\ 
        -\frac{k_m}{N}  & \mu_p+\mu_m & 0 & 0 &  &   \\ 
        0 & -\mu_m & \mu_p+\mu_m & 0 &  &   \\ 
        0 & 0 & -\mu_m & \mu_p+\mu_m &  &  \\ 
        \vdots &  &  & \ddots & \ddots &   \\ 
        0 &  &  &  & -\mu_m &  \mu_p+\mu_m \\         
    \end{bmatrix} 
    \overrightarrow{x}^{(1)} = \text{diag}(\overrightarrow{\kappa}) \overrightarrow{x}^{(0)}
\end{equation}
which gives the recursive relation
\begin{equation}
    x^{(1)}_q = \frac{\mu_m}{\mu_p + \mu_m} x^{(1)}_{q-1} + \frac{k_p(k_m/N)}{(\mu_p+\mu_m)(\mu_m+n(k_m/N))}
\end{equation}
where $x^{(1)}_q$ is the $q$-th element of $\overrightarrow{x}^{(1)}$. The characteristic equation of the recursion has the particular solution
\begin{equation}
    X=\frac{k_p}{\mu_p}\bigg[ \frac{k_m/N}{\mu_m+n(k_m/N)} \bigg] = \frac{\langle h \rangle}{n}. 
\end{equation}
Now, we can rewrite the recursive relation as
\begin{equation}
    x^{(1)}_q - \frac{\langle h \rangle}{n} = \frac{\mu_m}{\mu_p+\mu_m} \bigg( x^{(1)}_{q-1} - \frac{\langle h \rangle}{n} \bigg)
\end{equation}
which fully iterates as
\begin{equation}
    x^{(1)}_n - \frac{\langle h \rangle}{n} = \bigg(\frac{\mu_m}{\mu_p+\mu_m}\bigg)^{n-1} \bigg( x^{(1)}_{1} - \frac{\langle h \rangle}{n} \bigg).
\end{equation}
In combination {with the} first two rows of the matrix equation defined in (A4), we now have a closed system for $x^{(1)}_0$, $x^{(1)}_1$, and $x^{(1)}_n$. Solving the system, we find 
\begin{equation}
    x^{(1)}_1 = \frac{\langle h\rangle}{n} \bigg( 1- \frac{\mu_m\mu_p}{(\mu_p+\mu_m) \big( \mu_p+\frac{k_m}{N}(1-[\frac{\mu_m}{\mu_p+\mu_m}]^n) \big)} \bigg)
\end{equation}
and omit the solutions for $x_0^{(1)}$ and $x_n^{(1)}$ since they are not needed to proceed with the derivation. For convenience, let $y_n$ be the left-hand side of Eq. (A8). We can then rewrite Eq. (A8) as 
\begin{equation}
    y_n = \bigg( \frac{\mu_m}{\mu_p+\mu_m}\bigg)^{n-1} y_1
\end{equation}
where $y_q = x^{(1)}_q - \frac{\langle h \rangle}{n}$ for all $q$. Based on the repetitive structure of $\overrightarrow{\kappa}$, we can simply evaluate the inner product for the marginal second factorial moment {of} the protein count of the reduced model by summing over $x^{(1)}_q$ from $q=1$ to $q=n${; the sum} can be cleanly evaluated as a geometric series
\begin{equation}
    \langle h(h-1) \rangle = \frac{k_p}{\mu_p} \sum_{q=1}^n x^{(1)}_q = \frac{k_p}{\mu_p} \bigg( \sum_{q=1}^n \bigg[ \bigg( \frac{\mu_m}{\mu_m+\mu_p}\bigg)^{q-1} y_1 \bigg] + \langle h \rangle \bigg)  
\end{equation}
where the $\langle h \rangle$ term is to account for the fact that $y_q$ is shifted by $\frac{\langle h \rangle}{n}$ relative to $x^{(1)}_q$. With the second moment in hand, {using} Eq. (6) from the main text {and performing} a bit of algebra, $F$ simplifies to Eq. (39) of the main text.

\renewcommand{\theequation}{B\arabic{equation}}
\setcounter{equation}{0}
        
\section*{Appendix B: Fano Factor for transcriptional bursting under sRNA Regulation}
\subsection*{1. Mean}
To obtain the mean, let us first construct the transition rate matrix of the reduced model for the $b=1$ case
\begin{equation}
    T = 
     \begin{bmatrix}
        -\frac{k_m}{N} & \mu_1 & \mu_2 \\
        \frac{k_m}{N} & -\mu_1-\alpha & \beta \\
        0 & \alpha & -\mu_2-\beta
    \end{bmatrix}
\end{equation}
with the corresponding protein production rate vector $\overrightarrow{k}=(0,k_p,0)$. The nullspace of $T$ gives the stationary distribution 
\begin{align}
    \overrightarrow{x}^{(0)} = \frac{1}{A} \bigg({N (\beta \mu_1 + (\alpha + \mu_1) \mu_2)} , {k_m(\beta+\mu_2)},{k_m \alpha}\bigg)
\end{align}
with $A= N\beta \mu_1 + N (\alpha + \mu_1) \mu_2 + k_m (\alpha + \beta + \mu_2)$ to enforce probability conservation. From here, it follows that the mean $\langle H \rangle$ works out to be
\begin{align}
    \langle H \rangle &= \langle b \rangle \lim_{N \rightarrow \infty} N \langle h_1 \rangle \notag \\
    &=\lim_{N \rightarrow \infty} N \bigg[ \frac{\langle b\rangle}{\mu_p} (0,k_p,0) \cdot (p_0, p_1, p_2) \bigg] = \frac{\langle b \rangle k_mk_p(\beta+\mu_2)}{\mu_p(\beta \mu_1 + \mu_2(\alpha+\mu_1))}.
\end{align}

\subsection*{2. Variance \& Fano Factor}
To derive the variance, we need to get the second factorial moments of both the $b=1$ case and {the} $b=2$ case{,} for which {a separate kinetic scheme is} to be considered (Figure 4). For the second factorial moment of the $b=1$ case, we obtain the elements of $\overrightarrow{x}^{(1)}$ by solving the linear system
\begin{equation}
     \begin{bmatrix}
        \mu_p+\frac{k_m}{N} & -\mu_1 & -\mu_2 \\
        -\frac{k_m}{N} & \mu_p+\mu_1+\alpha & -\beta \\
        0 & -\alpha & \mu_p+\mu_2+\beta
    \end{bmatrix}    
    \overrightarrow{x}^{(1)} = \text{diag}\{{\overrightarrow{k}}\} \overrightarrow{x}^{(0)}
\end{equation}
where $j=1$ in Eq. (26) of the main text and $\overrightarrow{x}^{(0)}$ is given by Eq. (B2). Using the solution of $\overrightarrow{x}^{(1)}$, we can compute the second factorial moment for the $b=1$ case as
\begin{align}
    \langle h_1 (h_1-1) \rangle &= \frac{\overrightarrow{x}^{(1)}\cdot (0,k_p,0)}{\mu_p} \notag \\
    &={{k_m} {k^2_p} (\beta +{ \mu_2}) (\beta +{ \mu_2}+{ \mu_p}) ({k_m}+{ \mu_p} N)}
    \bigg/ 
    \bigg[{ \mu^2_p} ({k_m} (\alpha +\beta +{ \mu_2})+{ \mu_2} N (\alpha +{ \mu_1})+\beta  { \mu_1} N) \notag \\
    &\times({k_m} (\alpha +\beta +{ \mu_2}+{ \mu_p})+N ({ \mu_2}+{ \mu_p}) (\alpha +{ \mu_1}+{ \mu_p})+\beta  N ({ \mu_1}+{ \mu_p})) \bigg].
\end{align}
Now, we need to make use of the kinetic scheme depicted in Figure 4 for the $b=2$ case for which {we} define the transition rate matrix
\begin{equation}
    T_2 = 
    \begin{bmatrix}
        -\frac{k_m}{N} & 0 & \mu_1 & 0 & \mu_2 & 0 \\
        \frac{k_m}{N} & -2\mu_1-2\alpha & 0 & \beta & 0 & 0 \\
        0 & 2\mu_1 & -\mu_1 - \alpha & \mu_2 & \beta & 0 \\
        0 & 2\alpha & 0 & -\beta-\mu_2-\alpha-\mu_1 & 0 & 2\beta \\
        0 & 0 & \alpha & \mu_1 & -\mu_2-\beta & 2\mu_2 \\
        0 & 0 & 0 & \alpha & 0 & -2\beta-2\mu_2 \\
    \end{bmatrix}
\end{equation}
The corresponding production rate vector is $\overrightarrow{k}_2 = (0, 2k_p, k_p, k_p, 0, 0)$. Following {exactly} the \textit{same} {process for} $T_2$ as we did {for} $T$ {to obtain} the first two factorial moments, we have derived an expression for $\langle h_2(h_2-1) \rangle$ (too massive to be given here) that is used in conjunction with the expression for $\langle h_1(h_1-1) \rangle$ and $[\partial^2_z G_b(z)/\partial_z G_b(z)]\vert_{z=1} = 2 \langle b \rangle$ to obtain the variance in the protein distribution of the full model using Eq. (37) and then Eq. (6) of the main text. Dividing this by the result for the mean given in Eq. (B3), the {factors of} $N$ cancel and we arrive at the expression for the Fano factor of the protein distribution claimed in the main text.

\setcounter{equation}{0}
\renewcommand{\theequation}{C\arabic{equation}}
\section*{Appendix C: Exact Generating Function for the Stationary Protein Distribution of the nuclear export model}

In the reduced model, we are first interested in computing the probability distribution of the {waiting time} between consecutive protein arrival events in the Laplace domain. Since the mRNA stays in the cytoplasm immediately after a protein production event, $\phi(s)$ can be formulated as
\begin{equation}
    \phi(s) = \int_0^\infty k_p e^{-(\mu_m+k_p)t} e^{-st} \,dt + \psi(s) \bigg[\int_0^\infty \mu_m e^{-(\mu_m+k_p)t} e^{-st} \,dt\bigg] 
\end{equation}
where the first term {accounts for a} protein arrival occurring before mRNA degradation and the second term accounts for the case where the next protein arrival occurs after the reduced model reverts to the degraded mRNA state{,} so $\psi(s)$ is given by 
\begin{equation}
    \psi(s) = \bigg( \frac{k_m/N}{k_m/N + s}\bigg) \bigg( \frac{k_E}{k_E + s} \bigg) \phi(s).
\end{equation}
{Combining} equations (C1) and (C2) {and} solving for $\phi(s)$ yields
\begin{equation}
    \phi(s) = \frac{k_p(k_E+s)( \frac{k_m}{N}+s)}{(k_E+s)(\frac{k_m}{N}+s)(k_p+s) + s\mu_m(k_E+\frac{k_m}{N}+s)}.
\end{equation}
In the case where we do not allow for mRNA degradation by taking the limit $\mu_m \rightarrow 0$, we have a constitutively active mRNA strand and{,} correspondingly, $\phi(s)$ reduces to $k_p/(s+k_p)$ which is the Laplace transform of an exponential distribution as expected. Now, following Ref. \cite{Nossan-BJ-2024}, the stationary generating function of the $G/M/\infty$ queue takes the form of the generalized hypergeometric function
\begin{equation}
    \tilde{g}^\ast(z) = {_mF_n} \bigg[ \frac{a_1}{\mu_p}, \dots, \frac{a_m}{\mu_p};  \frac{b_1}{\mu_p}, \dots, \frac{b_n}{\mu_p}; c\mu_p^{m-n-1}(z-1) \bigg],
\end{equation}
and the set of parameters $\{ m, n, a_1, \dots, a_m, b_1, \dots, b_n \}$ {is} implicitly defined by
\begin{align}
    \frac{\phi(s)}{1-\phi(s)} &= \frac{c(s+a_1)\dots(s+a_m)}{s(s+b_1)\dots(s+b_n)}
\end{align}
where it is easy to see that $\phi(s)$ must be rational in order to parameterize $\tilde{g}^\ast(z)$. Using the result from Eq. (C3), we have that
\begin{equation}
    \frac{\phi(s)}{1-\phi(s)} = \frac{k_p(k_E+s)(k_m/N+s)}{s\big(s+ \frac{1}{2}(k_E + \frac{k_m}{N} + \mu_m + \sqrt{\Delta})\big) \big(s+ \frac{1}{2}(k_E + \frac{k_m}{N} + \mu_m - \sqrt{\Delta})\big)},
\end{equation}
where
\begin{align}
    \Delta &= \bigg( k_E - \frac{k_m}{N} \bigg)^2 - 2\mu_m\bigg( k_E + \frac{k_m}{N} \bigg) + \mu_m^2.
\end{align}
Putting it all together, the stationary generating function for the protein distribution in the reduced model is
\begin{equation}
    g^\ast(z) = {_2F_2}\bigg[ \frac{k_E}{\mu_p}, \frac{k_m/N}{\mu_p}; \frac{k_E + \frac{k_m}{N} + \mu_m + \sqrt{\Delta}}{2\mu_p}, \frac{k_E + \frac{k_m}{N} + \mu_m - \sqrt{\Delta}}{2\mu_p}; \frac{k_p}{\mu_p}(z-1) \bigg].
\end{equation}
Using Eq. (4), the stationary generating function for the original model claimed in the main text follows. 

\twocolumngrid

\end{document}